\documentclass[%
 aip,
rsi,%
 amsmath,amssymb,
 reprint,%
]{revtex4-1}
\usepackage{graphicx}
\usepackage{dcolumn}
\usepackage{amsmath}
\usepackage{bm}
\usepackage{color}
\usepackage{siunitx}
\usepackage{subfig}
\draft 

\begin{document}


\title{Considering non-uniform current distributions in magnetoresistive sensor designs and their implications for the resistance transfer function} 



\author{A. Bachleitner-Hofmann}
\author{C. Abert}
\affiliation{CD-Laboratory for Advanced Magnetic Sensing and Materials, University of Vienna, 1010 Vienna, Austria}
\author{H. Br\"uckl}
\affiliation{Center for Integrated Sensor Systems, Danube University Krems, 2700 Wiener Neustadt, Austria}
\author{A. Satz}
\affiliation{Infineon Technologies Austria AG, 9500 Villach, Austria}
\author{T. Wurft}
\author{W. Raberg}
\author{C. Pr\"ugl}
\affiliation{Infineon Techologies, 85579 Neubiberg, Germany}
\author{D. Suess,$^\text{1}$}
\noaffiliation


\date{\today}

\begin{abstract}
Non-uniform current distributions of spin valves with disk shaped free layers are investigated. In the context of spin valves, the vortex state, which is the ground-state in many disk shaped magnetic bodies, allows for distinct parallel channels of high and low resistivity. The readout current is thus able to evade high resistivity regions in favor of low resistivity regions, giving rise to 'conductive inhomogeneities'. Therefore, the total resistance of the spin valve does not always correspond exactly to the total average magnetization of the free layer. In addition, the resistance transfer function can be significantly influenced by the spatial placement of the electrodes, giving rise to 'geometric inhomogeneities'. The resulting deviations from resistance to magnetization transfer function are investigated for different spin valve geometries and compared to measurements of comparable devices.
\end{abstract}


\maketitle 

\section{Introduction}
\label{sec:introduction}
Magnetoresistive effects are increasingly becoming some of the most important magnetic sensor technologies for a wide variety of applications, including high volume fields like automotive and biomedical applications~\cite{freitas2003magnetoresistive}. However, when reading the output signal of a GMR/TMR spin valve, the directly measured quantity is not the magnetization, but the output voltage or current which usually is assumed to be directly proportional to the average free layer magnetization component in the sensitive direction. Depending on the sensor geometry and particularly for some magnetization configurations, the electrical output can deviate significantly from the value that is expected. The deviation stems from the presence of low- and high-resistance regions in the spin valve, since if the electrical current is given paths of different resistance from one electrical contact to the other, the path with lower resistance will be preferred. The low resistance regions are therefore given more weight in the output signal than their high resistance counterparts. The investigated free layer geometries are disk shaped, where the stable magnetization state, in the absence of an external field, is a vortex state. On one hand, the potentially large linear range and absence of magnetic hysteresis make spin valves with a vortex in the free layer promising candidates for new sensor designs~\cite{zimmer2013device}. On the other hand, a magnetic vortex state features large regions with coherent magnetization, presenting channels for both high and low resistance for readout currents, making such designs potentially susceptible to readout errors due to non-uniform current distributions. 

\begin{figure*}
	\centering
    \includegraphics[width=\textwidth]{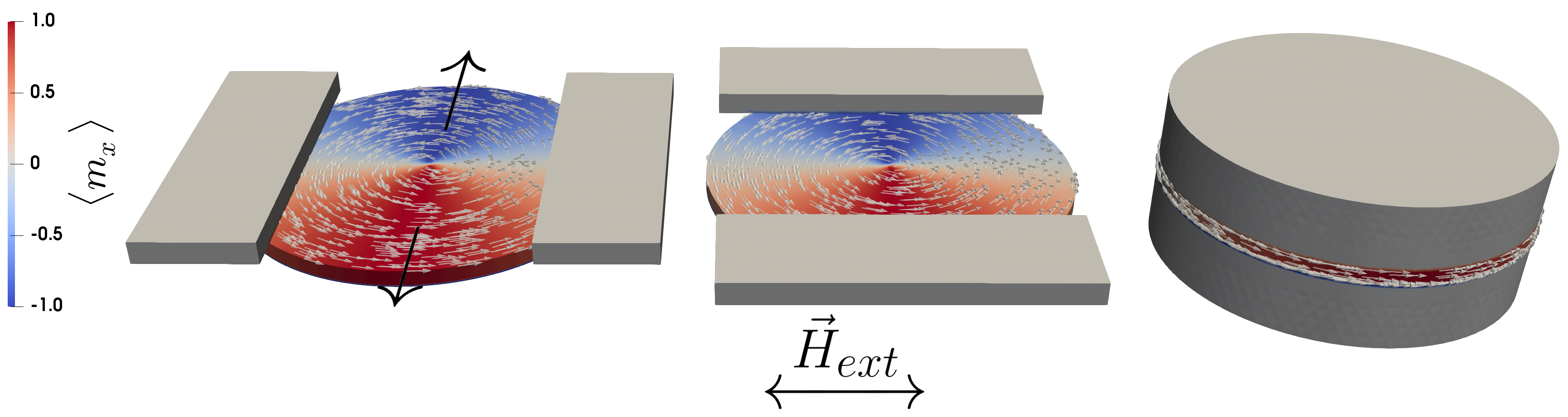}
    \caption{\label{fig:geometries}CIP and CPP spin valve geometries. The free layer is in a vortex state and the pinned layer is pinned in negative x direction, making the x-axis the sensitive field direction, while the axis of movement for the vortex core is the y-axis. Since the vortex has counter-clockwise helicity, the top half of the disk has lower, and the bottom half higher resistivity. In the CIP $\bm{j}\parallel m_\text{pinned}$ geometry (left), the readout current is presented channels of different resistivity, and will therefore increasingly prefer the lower resistivity channel the higher the MR ratio is. In the CIP $\bm{j}\perp m_\text{pinned}$ geometry (middle), the current is presented symmetric resistivity to either transverse direction, and is therefore much less able to evade high resistivity regions of the free layer. In the CPP geometry, similar to the $\bm{j}\parallel m_\text{pinned}$ case, the current is given paths of different resistivity.}
\end{figure*}

\section{Experimental details}
The magnetic hystereses of disks with diameter $d=\SI{1600}{\nano\meter}$ and thickness $t=\SI{65}{\nano\meter}$ were simulated using the finite elements package {\it femme}~\cite{suess2002time}. The micromagnetic simulation parameters were: saturation magnetization $\mu_0M_s=\SI{1.75}{\tesla}$, exchange stiffness $A=\SI[per-mode=fraction]{1.5e-11}{\joule\per\meter}$ and gilbert damping constant $\alpha=0.02$. The hysteresis loops were calculated using an LLG ramp method, were the Landau-Lifshitz-Gilbert equation is integrated over a long duration of time ($\SI{1.4}{\micro\second}$ per field period) during which the applied field is continually increased/decreased. The maximum applied fields were $\mu_0H_\text{max} = \SI{120}{\milli\tesla}$. For time integration, a higher order BDF scheme with adaptive time-stepping was used~\cite{suess2002time}, which usually employs timesteps in the order of picoseconds. Mesh sizes were approximately $\SI{12}{\nano\meter}$ for the micromagnetic simulation, and $\SI{5}{\nano\meter}$ (at the thin spacer layer) to $\SI{60}{\nano\meter}$ (at the far end of the electrodes) for the current distribution simulation.
The current paths for different electronic designs were calculated using the finite elements package {\it magnum.fe}~\cite{abert2013magnum}. To that end, simplified models consisting of free ($\SI{65}{\nano\meter}$), spacer ($\SI{2.5}{\nano\meter}$), and pinned layer ($\SI{5}{\nano\meter}$) as well as electrodes ($\SI{300}{\nano\meter}$) were created. The spacer layer and electrodes were assigned the conductivity of copper $\sigma_\text{Cu} = \SI[per-mode=fraction]{5.959e7}{\siemens\per\meter}$. The free and pinned layer were assigned the conductivity of amorphous Cobalt-Iron-Boron\cite{minor2003transverse} $\sigma_\text{CoFeB} = \SI[per-mode=fraction]{1.714e6}{\siemens\per\meter}$. 
For the TMR simulations, the effective conductivity, $\sigma_\text{MgO} = \SI[per-mode=fraction]{0.623}{\siemens\per\meter}$, for the MgO tunneling barrier was extracted from measurements of comparable TMR stacks, although the actual value is of little importance for our purposes, since the conductivity is several orders of magnitude lower than in the other layers and thus the current distribution is fully determined by the tunneling barrier anyway. The local resistivities/conductivities of the free layer were then adjusted in dependence of the local magnetization states from the micromagnetic simulation, according to simple models for the GMR and TMR effect respectively. The extent of the local resistivity changes was chosen so that the total magnetoresistive ratio of the spin valve, including electrodes, reflected measurements of comparable geometry~\footnote{All measurements were performed by Infineon Technologies}. For current-in-plane (CIP) GMR geometries, the measured effect was about $\SI{5}{\percent}$ at about $\SI{8.5}{\ohm}$. For current-perpendicular-to-plane (CPP) TMR geometries, the measured effect was about $\SI{55}{\percent}$ at about $\SI{0.8}{\kilo\ohm}$. For CPP GMR spin valves, no comparable devices were available. Nevertheless, in this paper we included CPP GMR devices using reported values for GMR ratios of up to $\SI{50}{\percent}$~\cite{bai2012data}.
\begin{figure}
	\centering
    \subfloat[\label{fig:CIP-measurement}]{\includegraphics[width=0.46\textwidth]{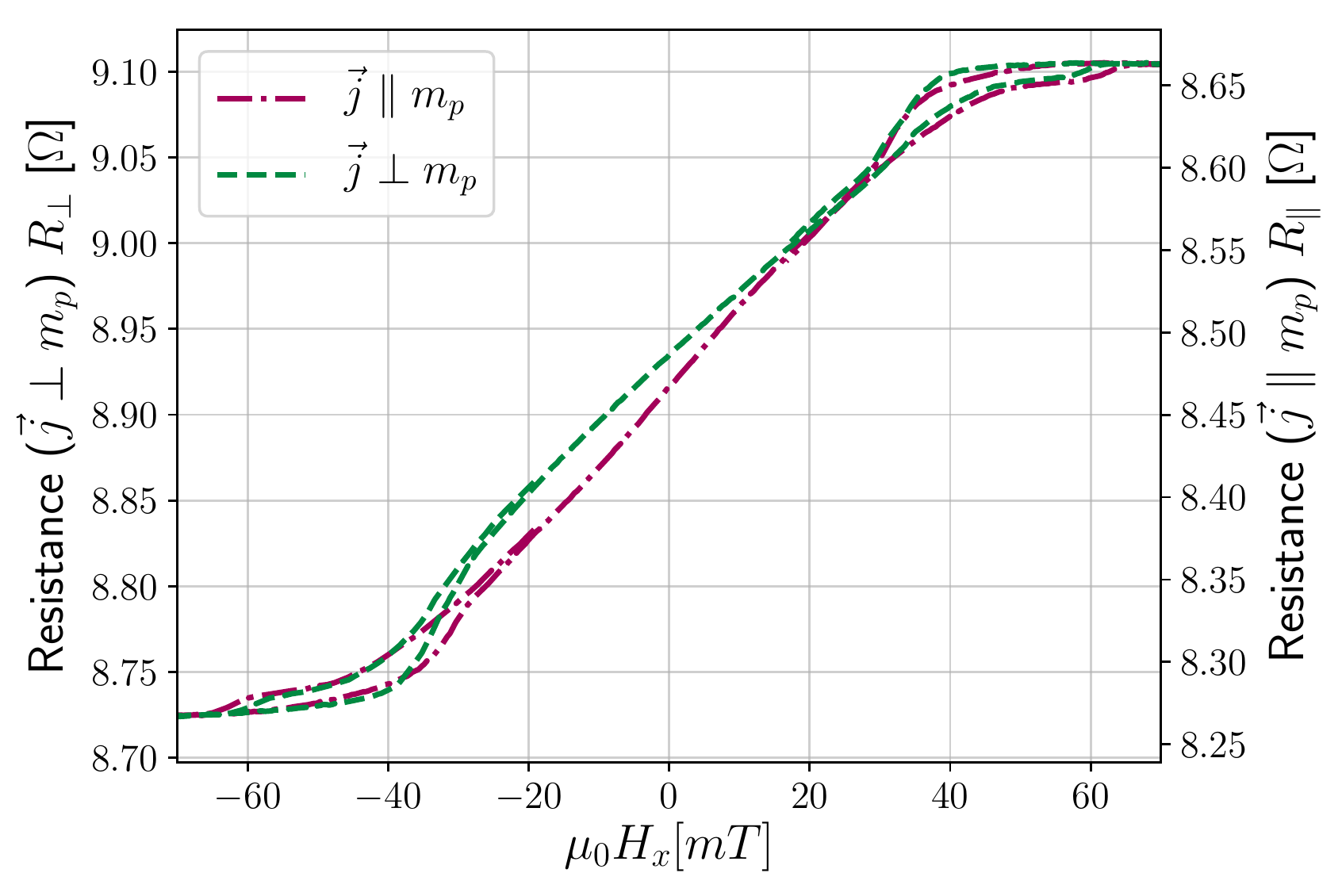}}\\
    \subfloat[\label{fig:CIP-simulation}]{\includegraphics[width=0.46\textwidth]{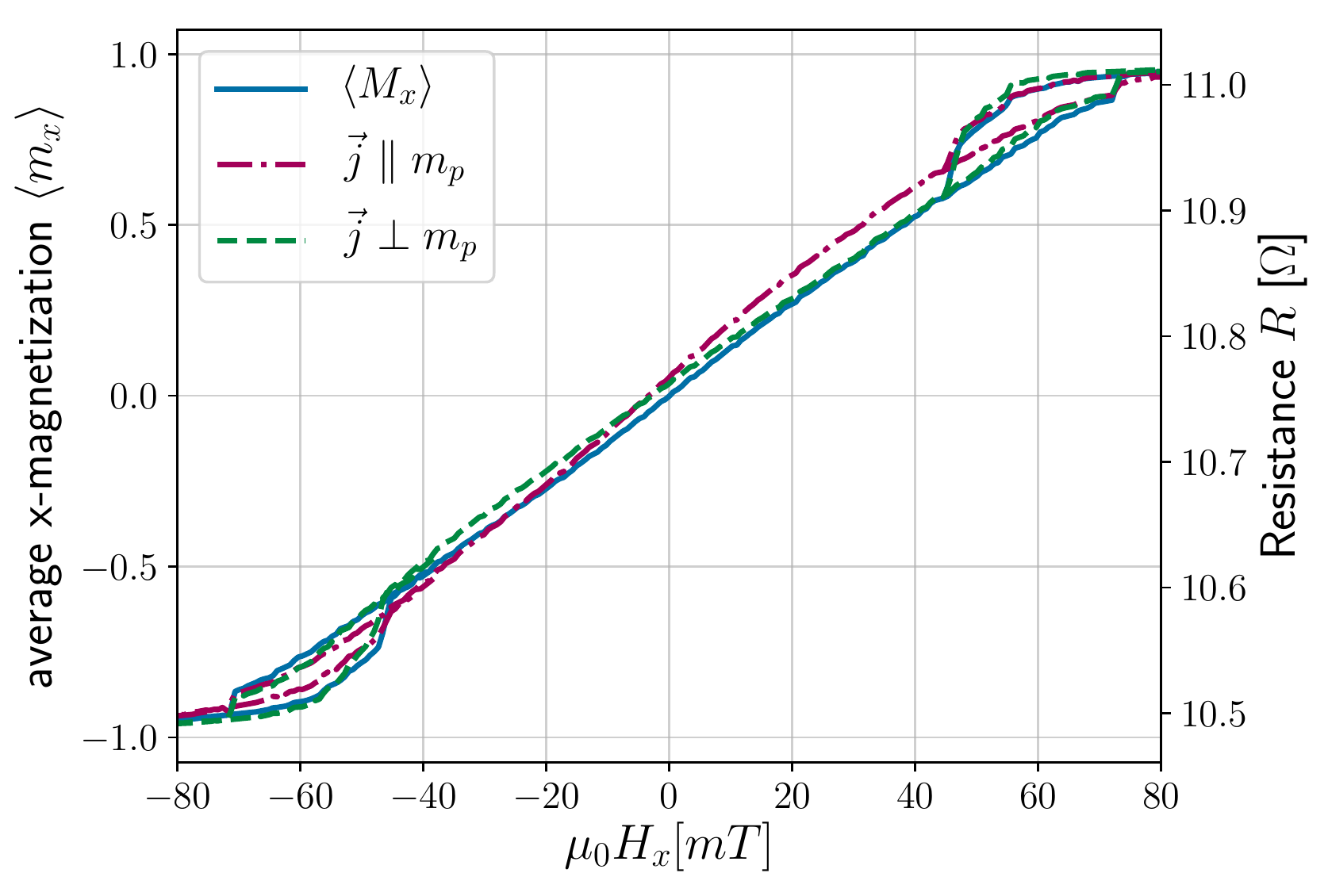}}
    \caption{\label{fig:CIP}Resistance over field of spin valves in $\vec{j}\parallel m_{pinned}$ and $\vec{j}\perp m_{pinned}$ current flow configurations. (a) measurement of stack with $\SI{2}{\micro\meter}\times\SI{80}{\nano\meter}$ CoFeB freelayer (b) Finite elements simulation of stack with $\SI{1600}{\nano\meter}\times\SI{65}{\nano\meter}$ CoFeB freelayer. The general trend of current distribution deviations could be reproduced by the simulations.}
\end{figure}
\section{CIP GMR}
\label{sec:CIP}
To determine local resistivities, we used a simple phenomenological model for the GMR effect
\begin{equation}
\label{eq:gmr}
\rho = \rho_\text{min}\left(1 + \delta_\text{GMR}\frac{1+\cos(\vartheta)}{2} \right)
\end{equation}
where $\delta_\text{GMR} = \frac{\rho_\text{max} - \rho_\text{min}}{\rho_\text{min}}$ is the local GMR ratio and $\vartheta$ is the angle between local free and pinned layer magnetization. 
The scaling length for the CIP GMR effect is the mean free path of the electrons~\cite{barthelemy2002magnetoresistance,fert2008nobel}, which is in the order of a few nanometers~\cite{gall2016electron} and which is dependent on crystalline structure and grain size~\cite{vancea1984mean,rubinstein1994classical}. Since the free layers of the structures investigated in this paper are considerably thicker than the mean free path, not all of the free layer contributes equally to the total magnetoresistance~\cite{dieny1992giant}. In our model, the local magnetoresistive effect was thus scaled by
\begin{equation}
\label{eq:exponentialdecayCIP}
\delta_\text{GMR} = \delta_\text{GMR,0} \cdot e^{-\frac{z}{\lambda_\text{mf}}}
\end{equation}
where $\delta_\text{GMR,0}$ is a constant related to the magnetoresistive effect, $z$ is the distance from the free/spacer layer interface, and $\lambda_\text{mf}$ is the electron mean free path in the free layer which was taken as $\SI{5}{\nano\meter}$. $\delta_\text{GMR,0}$ was chosen so that the total GMR effect of the spin valve $\frac{\Delta R}{R} = \frac{R_\text{max}-R_\text{min}}{R_\text{min}} = \SI{4.8}{\percent}$ matched the GMR effect of the measurements.

In the CIP geometry, the electrodes are typically placed on opposing sides of the spin valve. We can thus distinguish different cases of how the direction of the readout current is oriented relative to the sensitive axis of the spin valve (i.e. the pinned layer magnetization direction $\vec{m}_\text{p}$). In this work, the cases of parallel and perpendicular readout current were investigated (fig. \ref{fig:geometries} left and middle). For a magnetic vortex state in the free layer, these cases are quite different. If $\vec{j} \parallel \vec{m}_\text{p}$, the spin valve resembles a parallel circuit of a region with low resistivity and a region with high resistivity (fig. \ref{fig:geometries} left). If $\vec{j}\perp\vec{m}_\text{p}$, the spin valve resembles a series circuit of a region with low resistivity and a region with high resistivity (fig. \ref{fig:geometries} middle). The respective spatial ratio of low/high is determined by the vortex core position and thus by the external magnetic field. Fig. \ref{fig:CIP} shows measurement and simulation of comparable stacks in both configurations. In the linear range of $\mu_0H_\text{x} \approx \pm \SI{30}{\milli\tesla}$, the sensitivity of the parallel configuration is distinctly higher than in the perpendicular configuration, which could be reproduced by the simulation. The reason is, that because the shape of the free layer is a disk, and the electrodes only have a limited lateral contact area, more current is flowing through the center of the disk than on its sides. This means that in the parallel configuration, the spin valve is most sensitive when the vortex core is near the center, at low field amplitudes, and gets less sensitive as the vortex core approaches the disk edges, at high field amplitudes. In the perpendicular configuration, as the vortex core approaches the electrodes (see fig. \ref{fig:geometries} center), more current flows through the minority resistance region than is representative of the total spin valve state, leading to increased resistance at 'low resistance' states, and reduced resistance at 'high resistance' states.

\begin{figure}
	\centering
    \includegraphics[width=0.49\textwidth]{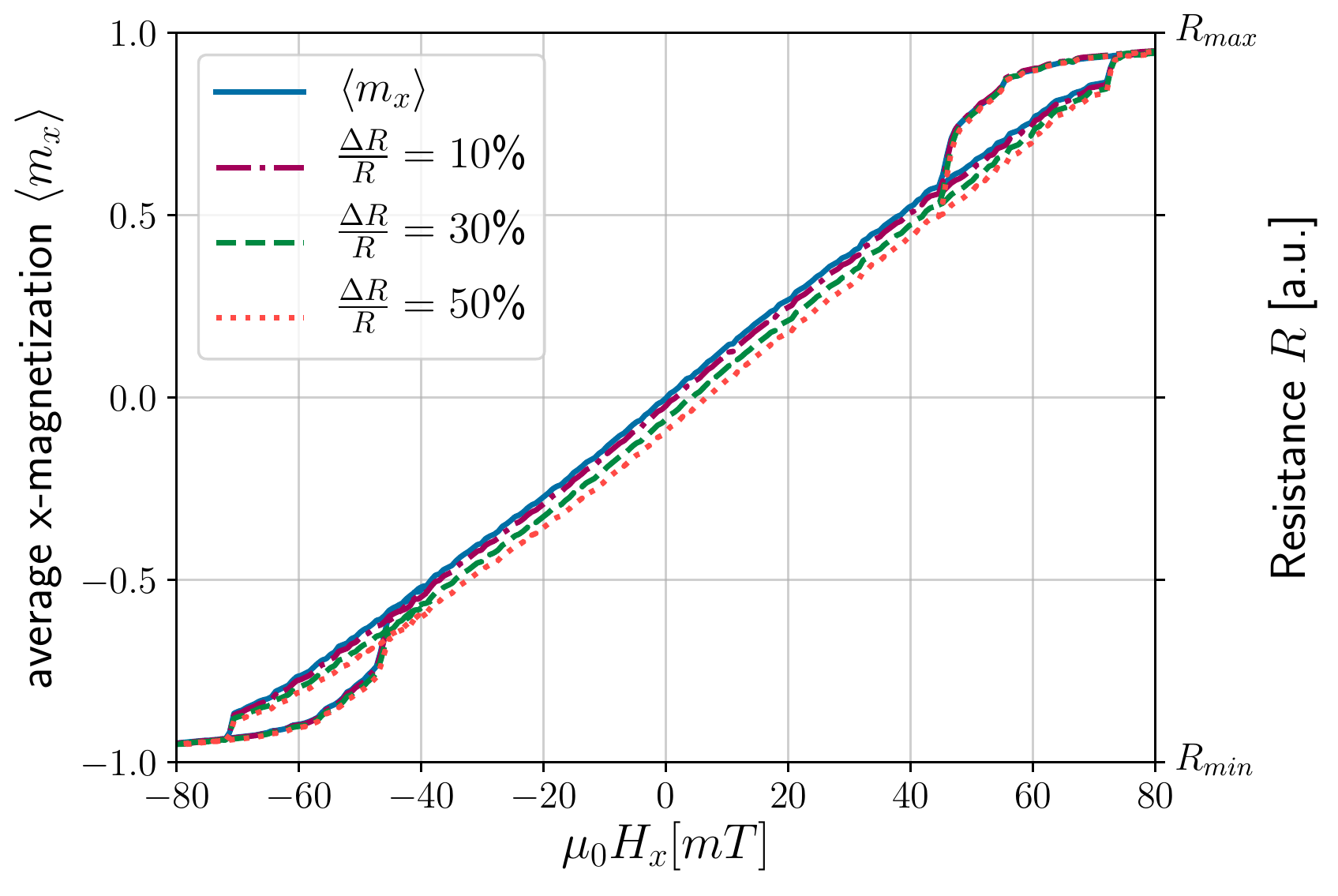}
    \caption{\label{fig:CPP}Simulation of a CPP GMR stack with a $\SI{1600}{\nano\meter}\times\SI{65}{\nano\meter}$ CoFeB freelayer. The electrodes were chosen to cover the full sensor area with $\SI{300}{\nano\meter}$ of Cu (fig. \protect{\ref{fig:geometries}} right). Due to higher thickness and higher conductivity in relation to the CoFeB free layer, the choice of $l_\text{sf}$ ($\widehat{=}$~active GMR region) did hardly influence the results because the current is given sufficient opportunity to adjust its path in the electrodes.}
\end{figure}
\section{CPP GMR}
\label{sec:CPP}
The scaling length of the CPP GMR effect is the spin diffusion length~\cite{barthelemy2002magnetoresistance}, which is usually much larger than the mean free path~\cite{fert2008nobel}. Similar to \eqref{eq:exponentialdecayCIP}, the local resistivity was scaled according to an exponential decay~\cite{valet1993theory}
\begin{equation}
\label{eq:exponentialdecayCIP}
\delta_\text{GMR} = \delta_\text{GMR,0} \cdot e^{-\frac{z}{l_\text{sf}}}
\end{equation}
where $l_\text{sf}$ is the spin-diffusion length. Although different spin diffusion lengths, from $\SI{5}{\nano\meter}$ to $\SI{65}{\nano\meter}$, have been investigated, the effects on the resistance transfer function were negligible. 
Since the electrodes are placed on top and bottom of the sensor stack, similar to the $\vec{j}\parallel\vec{m}_\text{pinned}$ CIP case, the sensor resembles a parallel circuit of high, intermediate and low resistivity regions. Fig. \ref{fig:CPP} shows simulations for $\SI{10}{\percent}$, $\SI{30}{\percent}$ and $\SI{50}{\percent}$ GMR effect at $l_\text{sf} = \SI{30}{\nano\meter}$. The resulting deviations from the magnetization hysteresis loop are a result of channels of different resistance, where the current flows preferentially through the lower resistance channels, giving more weight to corresponding magnetic moments towards the total output signal.

\section{CPP TMR}
\label{sec:TMR}
\begin{figure}
	\centering
    \subfloat[\label{fig:TMR-measurement}]{\includegraphics[width=0.49\textwidth]{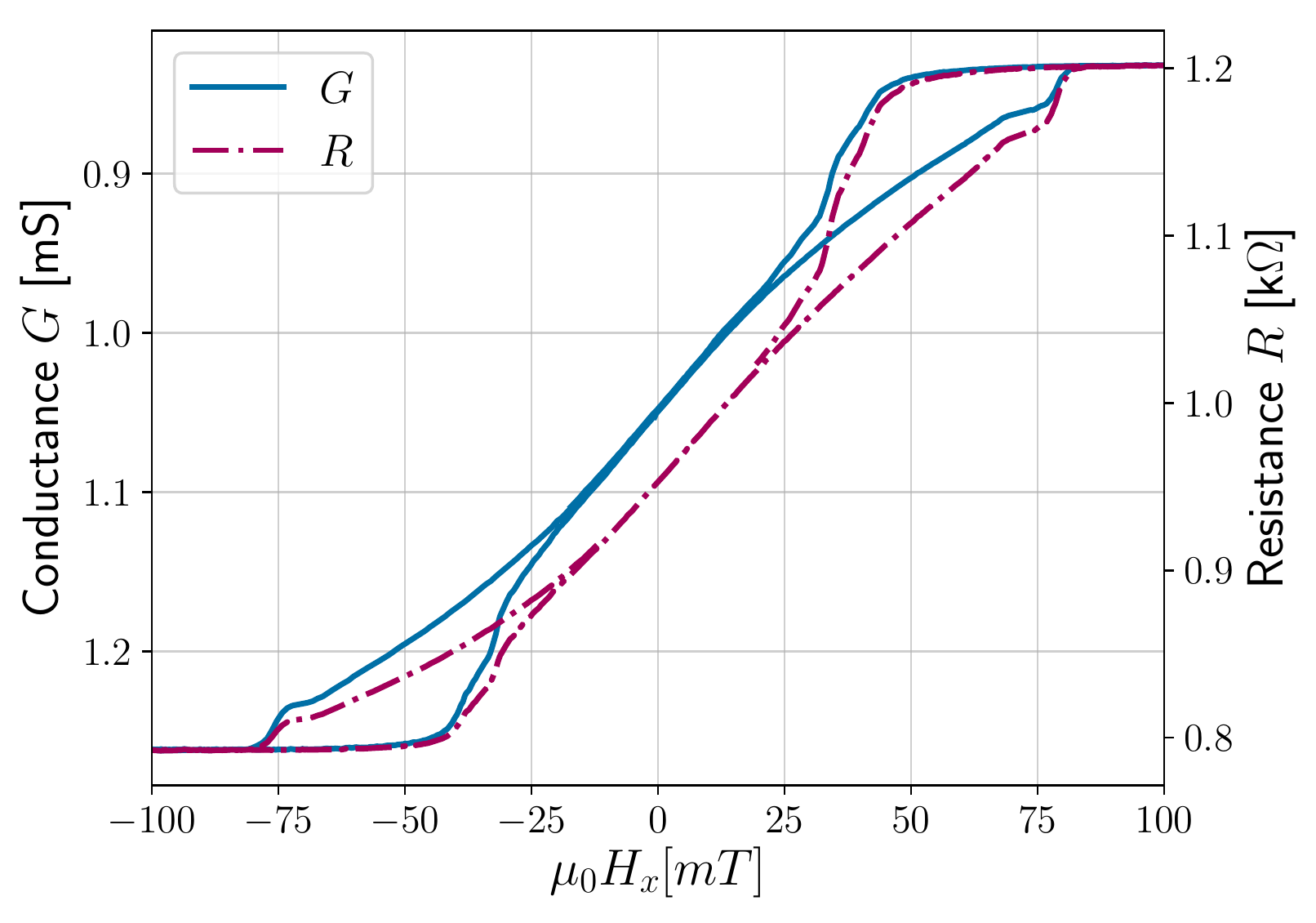}}\\
    \subfloat[\label{fig:TMR-simulation}]{\includegraphics[width=0.49\textwidth]{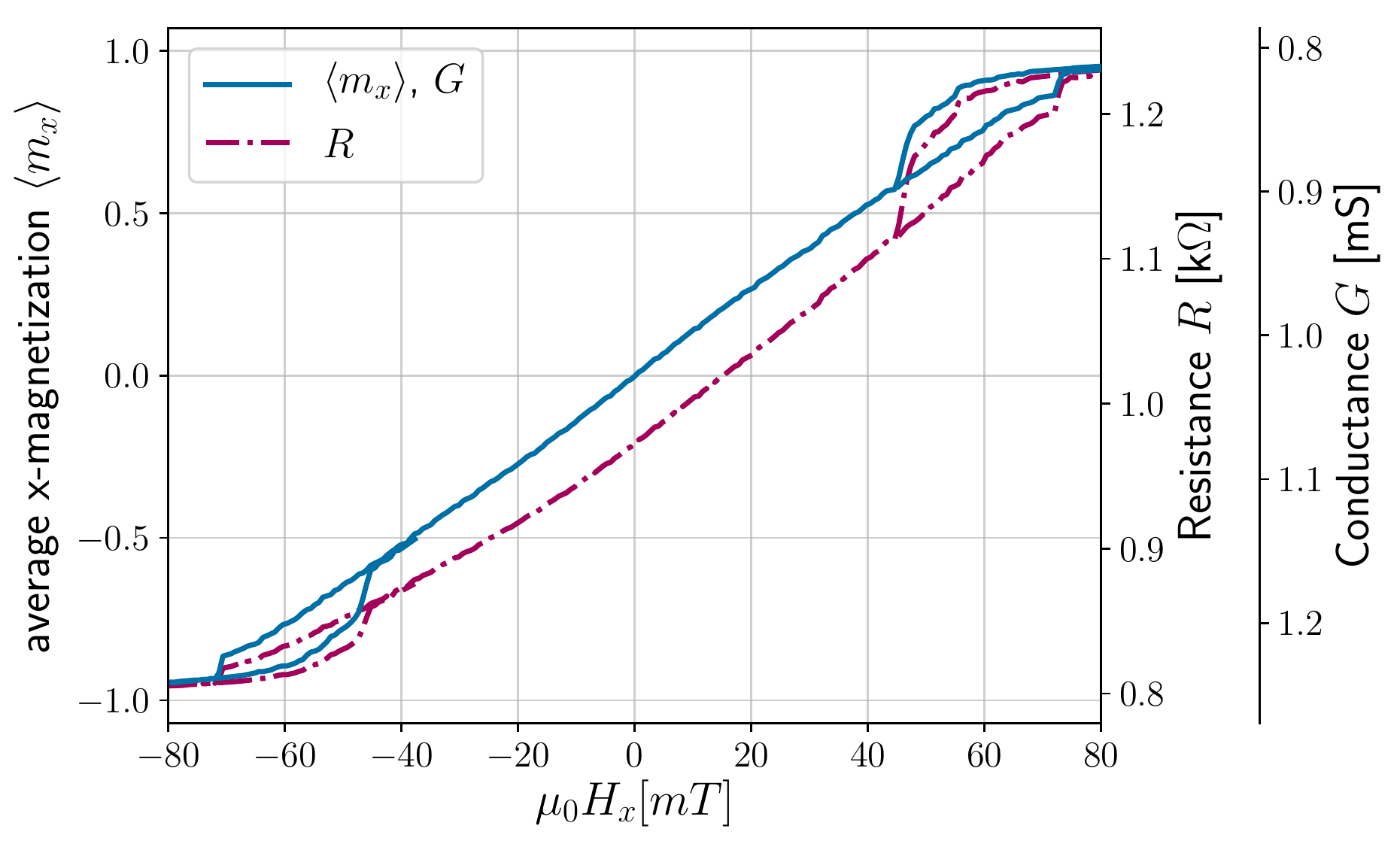}}
    \caption{\label{fig:TMR}(a) Measurement of a disk shaped TMR stack with an annihilation field of about $\SI{75}{\milli\tesla}$ (b) $\langle m_x(H_x) \rangle$, $R(H_x)$ and $G(H_x)$ curves for the finite elemets TMR current path simulation. Since the TMR effect is almost entirely originated in the very thin MgO layer, the spin valve resembles a perfect parallel circuit of different conductivities, all of which are linearly proportional to the respective local $m_x$. The total conductance is therefore linearly proportional to the average x-component of the magnetization and $G(H_x)$ is perfectly congruent with $\langle m_x(H_x) \rangle$ in the simulation.}
\end{figure}
In contrast to the GMR effect, where the resistivity is proportional to the dot product of local free and pinned layer magnetization, phenomenological models for the TMR effect employ proportionality of the conductivity instead of the resistivity~\cite{slonczewski1989conductance}. To model a magnetic tunneling junction (MTJ), a similar stack as in the CPP GMR simulations was used, except for a thicker pinned layer ($\SI{30}{\nano\meter}$) and a $\SI{1}{\nano\meter}$ MgO layer instead of the $\SI{2.5}{\nano\meter}$ Cu spacer layer. Since it is not directly related to the investigated current distributions, the well known dependence of the MTJ on the bias voltage~\cite{lu1998bias} was omitted in our simplified model. Analogously to \eqref{eq:gmr}, local conductivities of the MgO layer were determined by
\begin{equation}
\label{eq:tmr}
\sigma = \sigma_\text{min}\left(1 + \delta_\text{TMR}\frac{1+\cos(\vartheta)}{2} \right)
\end{equation}
where $\delta_\text{TMR} = \frac{\sigma_\text{max}-\sigma_\text{min}}{\sigma_\text{min}}$ is the local TMR ratio. Since the resistivity of the MgO tunneling barrier is by several orders of magnitude higher than the resistivity of CoFe and Cu, potential spin dependent scattering was neglected in these layers. Locally, the quantity proportional to the magnetization is the conductivity instead of the resistivity, the total average magnetization component in the sensitive direction and thus also the applied external magnetic field is therefore best represented by the total sensor conductance instead of the resistance. This has the advantage that in a parallel circuit the contribution of partial conductances toward the total conductance is linear, while for resistances it is not. Regardless of the extent of the TMR effect, the total conductance is therefore linearly proportional to the average magnetization and, by extension, also to the applied magnetic field. Fig. \ref{fig:TMR} shows the current path simulation for the $\SI{1600}{\nano\meter}\times\SI{65}{\nano\meter}$ free layer vortex stack and a measurement of a TMR stack with an amorphous disk shaped CoFeB free layer with similar annihilation field. In the simulation as well as in the measurement, we can see that the $G(H)$ curves are almost perfectly point-symmetric about their zero-field value, suggesting high congruence of $\langle m_\text{i}(H_\text{i}) \rangle$ and $G(H_i)$ curves, where i denotes the sensitive direction of the spin valve.

\section{Conclusion}
\label{sec:conclusion}
Our results show the influence of current distributions on the resistance/conductance transfer functions of GMR/TMR spin valves. In CIP GMR designs, the GMR effect is typically low at a few percent. The 'parallel circiut' current inhomogeneity is therefore of little consequence. Instead, the spatial placement of the electrodes and the subsequent accumulation of current on the central path through the disk leads to a 'geometric' current inhomogeneity which manifests differently for parallel and perpendicular configurations. \\
In GMR and TMR CPP stacks, the geometry is of little consequence to the current distributions, given that the electrodes and MR-inactive regions offer the current sufficient opportunity to select favorable paths. The 'parallel circuit' current inhomogeneity on the other hand leads to significant deviations of resistance to magnetization transfer function proportional to the MR effect. While this is a definite drawback of CPP GMR designs, for TMR stacks, the quantity proportional to the magnetization is the conductivity, and since in parallel circuits the total conductance is the sum of all conductances, the conductance of a TMR stack is perfectly proportional to the magnetization, regardless of the extent of the TMR effect.


%
%

%

\begin{acknowledgments}
The financial support by the Austrian Federal Ministry
of Science, Research and Economy, the National Foundation
for Research, Technology and Development and the Austrian
Science Fund (FWF): F4112 SFB ViCoM, is gratefully acknowledged.
\end{acknowledgments}

\nocite{*}
\bibliography{references}

\end{document}